\documentclass[12pt]{article}
\usepackage{latexsym}
\usepackage{epsfig}

\newcommand{\del}{\partial}
\newcommand{\beq}{\begin{eqnarray}}
\newcommand{\eeq}{\end{eqnarray}}
\newcommand{\be}{\begin{eqnarray*}}
\newcommand{\ee}{\end{eqnarray*}}

\newcommand{\bq}{{\bf q}}

\newcommand{\ra}{\rightarrow}
\newcommand{\e}{\epsilon}

\newcommand{\nn}{\nonumber}
\newcommand{\ket}[1]{\mbox{$\mid\!#1\rangle$}}
\newcommand{\bra}[1]{\mbox{$\langle#1\!\mid$}}

\begin{document}

\centerline{\large\bf {Graviton-photon conversion on spin 0 and 1/2 particles}}
\vskip 5mm
\centerline{Finn Ravndal and Mats Sundberg}
\vskip 5mm
\centerline{\it Institute of Physics, University of Oslo, N-0316 Oslo, Norway.}

\begin{abstract}

The differential cross-sections for scattering of gravitons into photons on bosons and fermions are
calculated in linearized quantum gravity. They are found to be strongly peaked in the forward direction and 
become constant at high energies. Numerically, they are very small as expected for such gravitational interactions.

\end{abstract}

\section{Introduction}
Inspired by string theories, it has recently been suggested that spacetime might have large, extra
dimensions so that gravity can be unified with the other interactions at experimentally
accessable energies\cite{ADD}\cite{RS}. As a consequence, the graviton will then have stronger
interactions than in more conventional theories. In addition there will be a large number of
Kaluza-Klein excitations of the graviton which also can be important in future 
experiments\cite{GRW}\cite{CPP}\cite{HLZ}\cite{Hewett}\cite{PerO}. These will have couplings of the same kind as the 
ordinary, massless graviton. For these reasons it will now be of some interest to study the interactions of 
gravitons with other particles in order to learn more about what to expect within such theoretical frameworks.

We will here calculate the differential cross-section for the graviproduction process $g + p \ra
\gamma + p'$ where a photon $\gamma$ is produced in the final state by an incoming graviton $g$ interacting
with a matter particle $p$. Using linearized quantum gravity and the requirements of gauge invariance,
the production amplitude is uniquely defined in lowest and dominant order of perturbation theory. In
the next section we give a short summary of linearized gravity and the Feynman rules for graviton
interactions. In Section 3 we calculate the production cross-section on scalar particles
and on Dirac particles in the following section. Finally, there is a short discussion of our results.

\section{Linearized theory of gravity}
In standard Einstein gravity, the metric $g_{\mu\nu}$ is determined by the energy-momentum tensor
$T_{\mu\nu}$ as the solution of the fundamental equation
\beq
      R_{\mu\nu} - {1\over 2}g_{\mu\nu}R = 8\pi G T_{\mu\nu}                     \label{Ein}
\eeq
where  $R_{\mu\nu}= R^\lambda_{\mu\lambda\nu}$ is the Ricci tensor and $R = R^\mu_{\;\;\mu}$. The constant 
multiplying $T_{\mu\nu}$ is the squared inverse Planck mass which we for convenience will denote by 
$f^2 = 8\pi G$. For small fluctuations around the Minkowski metric $\eta_{\mu\nu}$, the full metric is now 
usually written as\cite{Scadron}\cite{RPF}.
\beq
     g_{\mu\nu} = \eta_{\mu\nu} + 2fh_{\mu\nu}
\eeq
where $h_{\mu\nu}(x)$ is the graviton field. When it is small in amplitude, we then have to lowest order  
\beq
      R_{\mu\nu} = -f(\Box\,h_{\mu\nu} - \del_\mu h_\nu - \del_\nu h_\mu)       \label{Rnew}
\eeq
where $h_\mu = \del_\lambda h^\lambda_{\;\;\mu} - {1\over 2}\,\del_\mu h^\lambda_{\;\;\lambda}$. 
It is now easy to verify that this curvature tensor is invariant under the local gauge
transformation
\beq
     h_{\mu\nu} \ra h_{\mu\nu} + \del_\mu\chi_\nu  + \del_\nu\chi_\mu                 \label{gauge}
\eeq
where $\chi_\mu = \chi_\mu(x)$ is an arbitrary vector function. These gauge transformations correspond
to local coordinate transformations.

This local invariance allows us to choose a convenient gauge for the graviton field as one also does
in the description of the photon field. The simplest and most common gauge choice is the Hilbert or harmonic 
gauge defined by the condition $h_\mu = 0$. It is convenient to introduce the barred field
\beq
     \bar{h}_{\mu\nu} = h_{\mu\nu} - {1\over 2}\eta_{\mu\nu} h^\lambda_{\;\;\lambda}  \label{bar}
\eeq
which is an idempotent operation, i.e. $\bar{\bar{h}}_{\mu\nu} =  h_{\mu\nu}$. The Hilbert gauge condition 
can then be written as $\del_\mu\bar{h}^{\mu\nu} = 0$. The Ricci tensor (\ref{Rnew}) now 
takes the simple form $R_{\mu\nu} = -f\Box\,h_{\mu\nu}$ and the Einstein equation (\ref{Ein}) 
simplifies to
\beq
     \Box\,\bar{h}_{\mu\nu} = -fT_{\mu\nu}
\eeq
We have thus arrived at the wave equation for the graviton field produced by the source on the right-hand side.
Its stucture is the same as for the electromagnetic wave equation in the Lorentz gauge where the source is the
electric four-current.

Using the analogy with electromagnetic theory, we can now couple the above graviton field
to another matter distribution described by the energy-momentum tensor $T'_{\mu\nu}$ via the interaction
\beq
     {\cal L}_{int} = -f{h}^{\mu\nu}T'_{\mu\nu}                                       \label{Lint}
\eeq
This gives a direct coupling between these two matter distributions which is
\beq
     {\cal L}_{int} = f^2\,T'_{\mu\nu}{1\over\Box}\,\bar{T}^{\mu\nu}
\eeq
where the barred energy-momentum tensor $\bar{T}_{\mu\nu}$ is defined as in (\ref{bar}). It can be
un-barred with the help of the projection operator
\beq
     P_{\mu\nu\rho\sigma} = {1\over 2}(\eta_{\mu\rho}\eta_{\nu\sigma} + \eta_{\mu\sigma}\eta_{\nu\rho}
                          - \eta_{\mu\nu}\eta_{\rho\sigma})                              \label{P_mnrs}
\eeq
to give
\beq
     {\cal L}_{int} = f^2\,T'^{\mu\nu}{ P_{\mu\nu\rho\sigma}\over\Box}\,T^{\rho\sigma}
\eeq
In the quantized theory this interaction is due to the exchange of a graviton. We thus have for the
graviton propagator in momentum space
\beq
     D_{\mu\nu\rho\sigma}(k) =  - {P_{\mu\nu\rho\sigma}\over k^2 + i\e}
\eeq
From Eq. (\ref{Lint}) we see that it couples to the energy-momentum tensor with strength $f$. The
situation is very similar to QED where the photon field $A_\mu$ couples to the electrical current with 
strength $e$ which is the electric charge of the particle.

\subsection{Graviton polarizations}

The description of a free graviton is not completely fixed in the Hilbert gauge. It still allows
gauge transformations of the type (\ref{gauge}) as long as $\Box\chi_\mu = 0$. One can then impose further
conditions on the field tensor which now can be taken to be traceless, $h^\sigma_{\;\;\sigma} = 0$. Four more
degrees of freedom are removed and one is working in the transverse traceless (TT)
gauge\cite{MTW}. The number of independent field components is then $10 - 4 - 4 = 2$ corresponding to the two 
helicity states $\oplus$ and $\otimes$ of the graviton. As polarization basis states it is then convenient to take
\beq
    \epsilon^{\mu\nu}(\oplus) = \frac{1}{\sqrt{2}} \left( 
                             \begin{array}{llll}
                               0 & 0 & 0 & 0 \\
                               0 & 1 & 0 & 0 \\
                               0 & 0 & - 1 & 0 \\
                               0 & 0 & 0 & 0 \\
                             \end{array} 
                           \right)
  ,\;\;
    \epsilon^{\mu\nu}(\otimes) = \frac{1}{\sqrt{2}} \left( 
                            \begin{array}{llll}
                              0 & 0 & 0 & 0 \\
                              0 & 0 & 1 & 0 \\
                              0 & 1 & 0 & 0 \\
                              0 & 0 & 0 & 0 \\
                            \end{array} 
                          \right)                                         \label{grav.pol}
\eeq
for a graviton with four-momentum $q^\mu = (\omega_\bq,\bq)$ and energy $\omega_\bq = |\bq|$ where the 
three-momentum $\bq$ is along the $z$-axis. We see that the TT gauge conditions $q_\mu\epsilon^{\mu\nu} = 
\epsilon^\mu_{\;\;\mu} = 0$ are satisfied. With this choice one can then do polarization sums based on the
projection operator\cite{Scadron}
\beq
    \sum_{\lambda=\oplus,\otimes} \e_{ij}(\lambda) \e^\star_{lm}(\lambda) &=& 
      \frac{1}{2} {\Big [}  (\delta_{il}\delta_{jm}+\delta_{im}\delta_{jl} - \delta_{ij}\delta_{lm})
                     - \frac{\delta_{il}q_j q_m +\delta_{im}q_j q_l - \delta_{ij}q_l q_m}{\bq^2} \nn \\ 
                 &-& \frac{\delta_{jm}q_i q_l +\delta_{jl}q_i q_m - \delta_{lm}q_i q_j}{\bq^2}
                     + \frac{q_i q_j q_l q_m}{\bq^4} {\Big ]}                            \label{pol.sum_1}
\eeq
An alternative method is obtained by making use of gauge invariance. The summation over the two physical helicity
states can then be extended to include also the non-physical states which by themselves mutually cancel. Using
completeness of all these polarization states, we then have instead
\beq
    \sum_{\lambda=\oplus,\otimes} \e_{\mu\nu}(\lambda) \e^\star_{\rho\sigma}(\lambda) = P_{\mu\nu\rho\sigma} 
                                                                                      \label{pol.sum_2}
\eeq
which is the projection operator introduced in (\ref{P_mnrs}). In many instances this allows a simpler calculation.

\section{Transition vertex elements}

Lowest order matrix transition elements which describe the coupling of gravitons to other fields, are well known
and can be found in the literature\cite{Scadron}\cite{RPF}. But since we find a few minor, but crucial discrepancies
in these formulas, we will present corrected expressions here.

\subsection{Coupling to bosons}

Let us first consider the coupling of a charged boson with mass $m$ described by the complex Klein-Gordon field 
$\phi(x)$ to the electromagnetic field $A_\mu(x)$. Gauge invariance gives rise to the conserved electric current 
$J_\mu = i(\phi^*\del_\mu\phi - \phi\del_\mu\phi^*)$. The coupling of a photon with four-momentum $k^\mu$ and
polarization vector $\e^\mu$ is then given by the matrix element
\beq
     \e^\mu\bra{p'}J_\mu(0)\ket{p} = \e^\mu(p_\mu' + p_\mu)
\eeq
where $p_\mu$ and $p_\mu' = p_\mu + k_\mu$ are the initial and final momenta of the particle. Choosing to work
in the Lorentz gauge, we now find invariance of the matrix element under the gauge transformation $\e^\mu \ra \e^\mu 
+ a k^\mu$ where $a$ is some arbitrary constant.

The analogous result for the coupling of a graviton with four-momentum $q^\mu$ and polarization tensor $\e^{\mu\nu}$ 
is now seen from (\ref{Lint}) to be given by the corresponding matrix element $\bra{p'}T_{\mu\nu}(0)\ket{p}$ 
where $T_{\mu\nu}$ is the energy-momentum tensor. For the boson field under consideration it has the form
\beq
    T_{\mu\nu}^B = \del_\mu\phi^*\del_\nu\phi + \del_\mu\phi\del_\nu\phi^* 
               - \eta_{\mu\nu}(\del_\sigma\phi^*\del^\sigma\phi - m^2\phi^*\phi)           \label{T_munu}
\eeq 
\begin{figure}[htb]
  \begin{center}
    \epsfig{figure=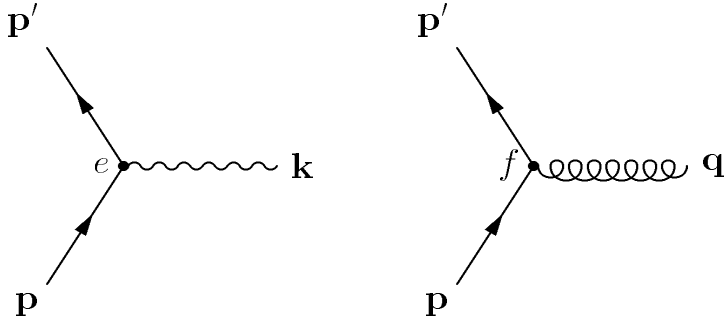,height=50mm,width=120mm}
  \end{center}
  \vspace{-10mm}                       
  \caption{\footnotesize Photon and graviton coupling to particle.}
\end{figure} 
Taking the matrix element between the same states, we find the gravitational coupling to a boson to be
\beq
    \e^{\mu\nu}\bra {p'} T_{\mu\nu}^B(0) \ket {p} 
  = \e^{\mu\nu}[p_\mu p'_\nu + p'_\mu p_\nu - \eta_{\mu\nu}(p \cdot p' -m^2)]
\eeq 
Gauge invariance is now verified by noting that under the substitution  $\e^{\mu\nu} \ra \e^{\mu\nu} +
q^\mu\chi^\nu + q^\nu\chi^\mu$ where $\chi_\mu$ is an arbitrary vector satisfying $q^\mu\chi_\mu = 0$, the coupling 
remains the same. These two couplings are shown in Fig.1.

There is also a contact term coupling a graviton directly to a photon in the presence of the particle field.
This follows from electromagnetic gauge invariance which implies that the partial derivatives $\del_\mu\phi$
in the energy-momentum tensor (\ref{T_munu}) must be replaced by the covariant derivatives 
$(\del_\mu  + ieA_\mu)\phi$. This generates new terms 
\beq
       T_{\mu\nu}^{AB} = & -& ie [(\phi^*\del_\nu\phi - \phi\del_\nu\phi^*)A_\mu 
                             + (\phi^*\del_\mu\phi - \phi\del_\mu\phi^*)A_\nu \nn \\
      &-&  \eta_{\mu\nu}(\phi^*\del_\lambda\phi - \phi\del_\lambda\phi^*)A^\lambda]  + {\cal O}(e^2)
\eeq
Taking the matrix element between an initial state containing a boson with momentum $p$ together with
a photon with momentum $k$ and polarization $\e^\lambda$ and a final state boson with momentum $p'$, 
we then find the matrix element
\beq
    \bra {p'} T_{\mu\nu}^{AB}(0) \ket {p;k,\lambda}   = - e [\eta_{\lambda\nu} (p_\mu + p'_\mu) 
  + \eta_{\lambda\mu} (p_\nu + p'_\nu) - \eta_{\mu\nu}(p_\lambda + p'_\lambda)]\e^\lambda    \label{T_AB}
\eeq      
\begin{figure}[htb]
  \begin{center}
    \epsfig{figure=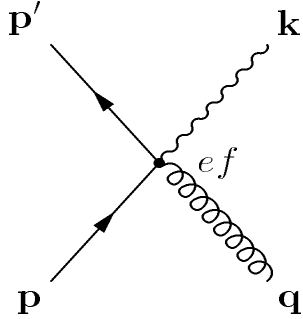,height=50mm,width=70mm}
  \end{center}
  \vspace{-10mm}            
  \caption{\footnotesize Graviton-photon contact interaction with particle.}
\end{figure} 
This vertex is not gauge invariant in itself, but must be combined with the previous couplings to give an overall
gauge invariant result. It is illustrated in Fig.2.

Gravitons couple to all kinds of matter and thus also to photons. From the electromagnetic energy-momentum tensor
\beq
      T_{\mu\nu}^{EM} = F_\mu^{\;\,\sigma} F_{\sigma\nu} +\frac{1}{4}\eta_{\mu\nu} F^{\rho\sigma} F_{\rho\sigma} 
\eeq
we can now find the matrix element for the transition between a photon with initial momentum $k$ and polarization 
$\e^\rho$ to a final momentum $k'$ and polarization $\e^\sigma$. We find the result
\beq
 \bra{k',\sigma} T_{\mu\nu}^{EM}(0)\ket{k,\rho} &=& \e^\sigma[k'_\rho (k_\mu\eta_{\sigma\nu} + k_\nu\eta_{\sigma\mu}) 
   + k_\sigma (k'_\mu\eta_{\rho\nu} + k'_\nu\eta_{\rho\mu}) \nn \\
   &-&  \eta_{\sigma\rho} (k'_\mu k_\nu + k'_\nu k_\mu) \nn +
    \eta_{\mu\nu} (\eta_{\sigma\rho}(k' \cdot k)  - k'_\rho k_\sigma) \nn \\ &-& 
    (k' \cdot k) (\eta_{\mu\rho} \eta_{\nu\sigma} + \eta_{\mu\sigma} \eta_{\nu\rho})]\e^\rho  \label{T_EM}
\eeq
which is in agreement with the literature\cite{Scadron} except for an overall factor of $1/2$. 
As the previous coupling of a graviton to a boson was shown in Fig.1, this new photon-graviton vertex is 
correspondingly illustrated in Fig.3.

\subsection{Coupling to fermions}
The electromagnetic coupling of a photon to a Dirac particle described by the free Lagrangian ${\cal L}_F 
= \bar{\psi}(i\gamma^\mu\del_\mu - m)\psi$ is via the conserved current $J^\mu = \bar{\psi}\gamma^\mu\psi$. 
The  matrix element for such a transition from an initial fermion state described by the Dirac spinor $u(p)$ 
where $p$ is the particle four-momentum, to a final fermion state $u(p')$, is then the well-known result 
$\bra{p'}J^\mu(0)\ket{p} = \bar{u}(p')\gamma^\mu u(p)$.
\begin{figure}[htb]
  \begin{center}
    \epsfig{figure=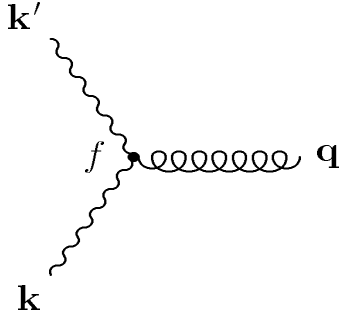,height=50mm,width=65mm}
  \end{center}
  \vspace{-10mm}            
  \caption{\footnotesize Graviton coupling to photon.}
\end{figure}

In order to find the energy-momentum tensor for a Dirac particle, we must make the above Lagrangian 
symmetric in ingoing and outgoing fields. It then takes the
form
\beq
    {\cal L}_F = {1\over 2}\bar{\psi}i\gamma^\mu(\del_\mu\psi) - {1\over 2}(\del_\mu\bar{\psi})i\gamma^\mu\psi
             - m\bar{\psi}\psi
\eeq
The energy-momentum tensor follows by standard methods. It allows the calculation of the matrix element
for a transition from an initial fermion state with four-momentum $p$ described by the Dirac spinor $u(p)$, 
to a final fermion state $u(p')$. We then obtain
\beq
        \bra {p'} T_{\mu\nu}^F(0) \ket {p} = {1\over 4}\bar{u}(p')[ (p_\mu + p'_\mu)\gamma_\nu 
                     + (p_\nu + p'_\nu)\gamma_\mu  - 2\eta_{\mu\nu}(\not \! p \,\; + \not \! p'-2m)]u(p)
\eeq
where we use the standard notation $\not \!\! p\equiv \gamma_\mu p^\mu$. It has the same form as for the coupling
of a graviton to a boson in Fig.1. We notice that the last term in this fermionic vertex is zero when the
initial and final particles are on-shell. This explains why it has been left out in the literature\cite{Scadron}.

There is also a graviton-photon contact term here. As for bosons, it can be found from the minimal
substitution $\del_\mu \ra \del_\mu  + ieA_\mu$ in the energy-momentum tensor. The resulting four-particle vertex is
then
\beq
         \bra {p'} T_{\mu\nu}^{AF}(0) \ket {p;k,\lambda} = -{e\over 2}\bar{u}(p')[\eta_{\mu\lambda} \gamma_\nu 
               +  \eta_{\nu\lambda} \gamma_\mu  -2\eta_{\mu\nu}\gamma_\lambda]u(p) \e^\lambda
\eeq
in analogy with (\ref{T_AB}) for bosons. It can also be illustrated as in Fig.2.

\section{Differential cross-sections}

We will now calculate the lowest order scattering amplitude for the inelastic process 
$g(q) + p \ra \gamma(k) + p'$ where a graviton
with momentum $q$ scatters on a particle with momentum $p$ such that it is converted into a photon with
momentum $k$ and a final particle with momentum $p'$. The matter particle has the mass $m$. Since both the 
graviton and photon are massless, the kinematics will be the same as for Compton scattering. In the laboratory 
system where the initial particle is at rest and the incoming graviton has energy $\omega$, the final state photon
will come out at an angle $\theta$ and with energy $\omega'$ given by the Compton formula
\beq
     {1\over\omega'} = {1\over\omega} + {1\over m}(1 - \cos\theta)
\eeq
We will assume that all particles are unpolarized. 

From the previous couplings, we find four diagrams which contribute to the process in lowest order. 
\begin{figure}[htb]
  \begin{center}
    \epsfig{figure=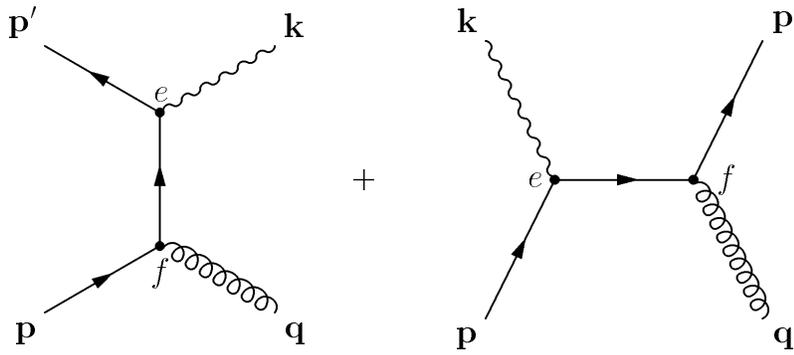,height=60mm,width=130mm}
  \end{center}
  \vspace{-10mm}            
  \caption{\footnotesize Direct and exchange production of photon by incoming graviton.}
\end{figure}
In addition to the two ordinary contributions shown in Fig.4, there is the contact diagram in Fig.2 and also the 
diagram in Fig.5 where the incoming graviton couples to an exchanged photon.

\subsection{Bosonic conversion}

When the graviton-photon conversion takes place on a boson, the first diagram in Fig.4 gives rise to the
transition matrix element
\beq
    M_{fi}^B(1) = - \frac{ef}{2p \cdot q}\e^{\mu\nu}[2p_\mu p_\nu + p_\mu q_\nu + p_\nu q_\mu - \eta_{\mu\nu}(q\cdot p)]
                (2p'_\lambda + k_\lambda)\e^\lambda  
\eeq
The contribution from the second diagram can similarly be written as
\beq
 M_{fi}^B(2) = \frac{ef}{2p \cdot k}\e^{\mu\nu}[2p_\mu' p_\nu' - p_\mu' q_\nu - p_\nu' q_\mu + \eta_{\mu\nu}(q\cdot p')]
                (2p_\lambda - k_\lambda)\e^\lambda  
\eeq
There is no internal propagator in the contact term diagram Fig.2  and its contribution to the scattering
amplitude can be read off directly from the bosonic vertex (\ref{T_AB}) and leads to
\beq
    M_{fi}^B(3) = ef \e^{\mu\nu}[\eta_{\lambda\nu} (p_\mu + p'_\mu) 
  + \eta_{\lambda\mu} (p_\nu + p'_\nu) - \eta_{\mu\nu}(p_\lambda + p'_\lambda)]\e^\lambda   
\eeq 
\begin{figure}[htb]
  \begin{center}
    \epsfig{figure=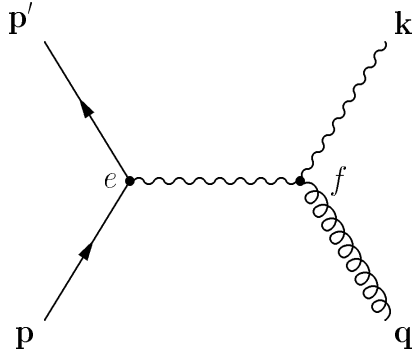,height=60mm,width=90mm}
  \end{center}
  \vspace{-10mm}            
  \caption{\footnotesize Virtual photon becomes real in the interaction with incoming graviton.}
\end{figure}
In the final diagram there is a virtual photon exchanged between the incoming graviton and the outgoing photon.
With the corresponding vertex from (\ref{T_EM}), we find the amplitude
\beq
    M_{fi}^B(4)  &=& - \frac{ef}{2 k \cdot q } \e^{\mu\nu} (p'^\rho + p^\rho)
                   [(k_\mu - q_\mu)(k_\rho \eta_{\lambda\nu} - k_\nu\eta_{\rho\lambda}) \nn \\
                      & + & (k_\nu - q_\nu)(k_\rho \eta_{\lambda\mu} - k_\mu\eta_{\rho\lambda})  
                   + (k_\lambda - q_\lambda)(k_\mu \eta_{\nu\rho} + k_\nu \eta_{\mu\rho} - k_\rho\eta_{\mu\nu}) \nn \\
                      &+& (k\cdot q)(\eta_{\mu\rho}\eta_{\nu\lambda} + \eta_{\nu\rho}\eta_{\mu\lambda} 
                     - \eta_{\mu\nu}\eta_{\rho\lambda})]\e^\lambda
\eeq 

The full scattering amplitude is now
\beq
    M_{fi}^B =  M_{fi}^B(1)+  M_{fi}^B(2)+ M_{fi}^B(3)+ M_{fi}^B(4)
\eeq
We check it for electromagnetic gauge invariance in the Lorentz gauge where $\e^\lambda k_\lambda = 0$ by verifying 
that it is 
invariant under the gauge transformation $\e^\lambda \ra \e^\lambda + a k^\lambda$ where $a$ is an arbitrary
constant. Similarly, we verify that it is invariant under the gravitational gauge transformation $\e^{\mu\nu} 
\ra \e^{\mu\nu} + q^\mu\chi^\nu + q^\nu\chi^\mu$ where $\chi_\mu$ is an arbitrary vector satisfying $q^\mu\chi_\mu = 0$
in the TT gauge where $\e^{\mu\nu}q_\mu = 0$ in momentum space\cite{Mats}.

The differential cross-section for the process $g + p \ra \gamma + p'$ in the laboratory system is now given by
\beq
    \frac{d\sigma}{d\Omega}{\Big |}_{lab} = \frac{1}{64\pi^2 m^2}\Big(\frac {\omega'}{\omega}\Big)^2 |M_{fi}|^2 
\eeq
when averaged over initial and summed over final spins. Averaging over the two initial graviton helicities
can be done either directly from the TT polarization tensors (\ref{grav.pol}) or using the corresponding
polarization projection operator (\ref{pol.sum_1}). Either way gives the same result
\beq
         |M_{fi}|^2 = e^2f^2m^2 (1+\cos^2\theta)\cot^2\frac{\theta}{2}               \label{lab}
\eeq 
after also summing over the two final photon helicities\cite{Mats}. It gives the cross-section
\beq
       \frac{d\sigma}{d\Omega}{\Big |}_{lab} = \frac{\alpha G}{2}
       \Big ( \frac {\omega'}{\omega} \Big )^2 (1+\cos^2\theta) \cot^2\frac{\theta}{2}
\eeq 
where $\alpha = e^2/4\pi$ is the fine-structure constant and $G = f^2/8\pi$ is Newton's gravitational
constant. The factor $(1+\cos^2\theta)$ is the same as for photon 
Compton scattering on a scalar particle, while the extra factor $\cot^2(\theta/2)$ makes the differential 
cross-section diverge in the forward direction. This is due to the exchange of the massless photon in Fig.5.
Since the masses of the particles in the initial and final states are the same, this will also be the cross-section
for the inverse process $\gamma + p \ra g + p'$ which is photoproduction of gravitons.

As an independent check of this surprisingly simple result, we have also evaluated the averaged square matrix
element in an arbitrary frame using the covariant projection operator $(\ref{pol.sum_2})$ for the graviton
and the corresponding result $\sum_\e\e_\mu\e_\nu = -\eta_{\mu\nu}$ for the photon. Introducing the
Mandelstam variables $s = (q + p)^2, t = (k - q)^2$ and $u = (p' - q)^2$ with $s + u + t = 2m^2$, we then find
\beq
|M_{fi}|^2  &=& e^2f^2\left[\frac{(u-m^2)(s-m^2)}{t} + m^2\Big ( 2+ \frac{(s-m^2)^2}{(u-m^2)^2}\right. \nn \\ 
     &+ & \left.2t \frac{(s+m^2)}{(u-m^2)^2} + t^2\frac{(s+m^2)^2-2m^4}{(u-m^2)^2(s-m^2)^2} \Big ) \right] \label{cov}
\eeq
When this is now evaluated in the laboratory system where
\beq
    s &=& m^2 + 2 m \omega      \nn \\ 
    t &=& -2\omega\omega'(1-\cos \theta)  \nn \\   
    u &=& m^2 - 2m\omega'                              \label{lab.sys}
\eeq
we find that (\ref{cov}) simplifies to the previous result (\ref{lab}).

The covariant result (\ref{cov}) also allows us to find the cross-section in the center-of-momentum frame.
In the high-energy limit where we can neglect the mass $m$, it becomes
\beq
    \frac{d\sigma}{d\Omega}{\Big |}_{CM} = \frac{\alpha G}{2} \cot^2\frac{\theta_{CM}}{2} 
\eeq
and shows the same forward peaking. It is independent of energy since it is due to the spin-1 photon
exchange that dominates at high energies.

\subsection{Fermionic conversion}

The Feynman diagrams that contribute to the process when the target particle is a fermion, are the same as in
the previous bosonic case. We thus have again four partial amplitudes which we label correspondingly. 
From the previously derived vertices, they are found to be
\beq
    M_{fi}^F(1) &=& - \frac{ef}{8p\cdot q}  \epsilon^{\mu\nu}
                     \bar{u}(p') \gamma_\lambda (\not \! p \, + \not \! q \, + m)
                       [ (2p_\mu + q_\mu) \gamma_\nu \nn \\ &+& (2p_\nu + q_\nu) \gamma_\mu 
                       -2\eta_{\mu\nu}(2\!\! \not \! p \, +\not \! q -2m) ] u(p) \e^\lambda   \\
    M_{fi}^F(2) &=&   \frac{ef}{8p\cdot k} \epsilon^{\mu\nu}
                      \bar{u}(p') [ (2p'_\mu - q_\mu) \gamma_\nu + (2p'_\nu - q_\nu) \gamma_\mu \nn \\ 
                      &-&2 \eta_{\mu\nu} (2\!\!\not \! p' - \not \! q - 2m) ]  
                      (\not \! p \, - \not \! k + m) \gamma_\lambda u(p)\e^\lambda  \\ 
    M_{fi}^F(3) &=&   \frac{ef}{2} \epsilon^{\mu\nu} \bar{u}(p') 
                      [\eta_{\lambda\mu}\gamma_\nu + \eta_{\lambda\nu}\gamma_\mu 
                     -2\eta_{\mu\nu} \gamma_\lambda]u(p)\e^\lambda   \\
    M_{fi}^F(4) &=&  - \frac{ef}{2 k \cdot q} \epsilon^{\mu\nu} \bar{u}(p')\gamma^\rho u(p,s) 
                       [(k_\mu - q_\mu)(k_\rho \eta_{\lambda\nu} - k_\nu\eta_{\rho\lambda}) \nn \\
                      & + & (k_\nu - q_\nu)(k_\rho \eta_{\lambda\mu} - k_\mu\eta_{\rho\lambda})  
                   + (k_\lambda - q_\lambda)(k_\mu \eta_{\nu\rho} + k_\nu \eta_{\mu\rho} - k_\rho\eta_{\mu\nu}) \nn \\
                      &+& (k\cdot q)(\eta_{\mu\rho}\eta_{\nu\lambda} + \eta_{\nu\rho}\eta_{\mu\lambda} 
                     - \eta_{\mu\nu}\eta_{\rho\lambda})]\e^\lambda
\eeq
The full scattering amplitude  $M_{fi}^F =  M_{fi}^F(1)+  M_{fi}^F(2)+ M_{fi}^F(3)+ M_{fi}^F(4)$ is again found to be
invariant both under electromagnetic and gravitational gauge trans\-formations. Summing over final and
averaging over initial spins in the squared amplitude, we find
\beq
      |M_{fi}|^2  &=&- e^2f^2\left[ 2s +t +2\frac{s^2}{t} + m^2 \Big(\frac{t}{s-m^2} - \frac{2s}{t}\right.  \nn \\ 
                  &+& \left.\frac{t(5s +3m^2 -t)}{(u-m^2)^2} - \frac{2(s-m^2)^3}{t(u-m^2)^2}
                   + \frac{4s^2t^2}{(s-m^2)^2(u-m^2)^2} \Big) \right] 
\eeq 
In the laboratory system where the Mandelstam variables take the values (\ref{lab.sys}), we then obtain the
cross-section
\beq
     \frac{d\sigma}{d\Omega}{\Big |}_{lab} = \frac{\alpha G}{2}\Big ( \frac {\omega'}{\omega} \Big )^2 
         \left[ \frac{\omega \omega'}{m^2} \sin^2 \theta + (1+\cos^2\theta) \cot^2\frac{\theta}{2} \right]  
\eeq
We see that the last term is the same as for scattering on a boson while the first term is due to the spin
of the target particle. 

In the center-of-momentum system the differential cross-section follows directly from
the above squared amplitude. At sufficiently high energies we can ignore the mass $m$ of the target particle
and then obtain
\beq
    \frac{d\sigma}{d\Omega}{\Big |}_{CM} = \alpha G \left[{1\over \sin^2(\theta_{CM}/2)} 
                                         - \frac{1}{2}\Big(1+\cos^2{\theta_{CM}\over 2}\Big) \right]
\eeq
In this limit we thus find again a cross-section which is independent of the energy and strongly peaked in the forward
direction.

\section{Concluding comments}

The typical size of these cross-sections is set by Newton's gravitational coupling constant which in our units
is $G = 2.61\times 10^{-66}$\,cm$^{2}$. Graviton-induced creation of photons or photoproduction of gravitons
will therefore be completely negligible under ordinary conditions. The only circumstances which could make these
reactions physically relevant, would be when the flux of incoming particles became extremely high so to
compensate for the very small coupling constant. A directed pulse of gravitons could then be converted into a
highly collimated beam of photons. It is not impossible to imagine such a process taking place in the
environment of highly compact and strongly relativistic astrophysical objects.

We want to thank R. Madsen for several useful discussions about linearized quantum gravity and A. Hiorth for
help with the figures.

\end{document}